# Images encryption using AES and variable permutations

V. M. Silva-García

Instituto Politécnco Nacional, CIDETEC, México.

Tel: 52-015-5557296000 ext. 52536     E-mail: vmsg@ipn.mx

R. Flores-Carapia     (Corresponding author)

Instituto Politécnco Nacional, CIDETEC, México.

Tel: 52-015-5557296000 ext. 52536     E-mail: rfloresca@ipn.mx

C. Rentería-Márquez

Instituto Politécnco Nacional, ESFM, México.

E-mail: crenteriam@gmail.com

B. Luna-Benoso

Instituto Politécnco Nacional, ESCOM, México.

E-mail: mobius_95@hotmail.com

C. A. Jiménez Vázquez

Instituto Politécnco Nacional, CIDETEC, México.

E-mail: cesar.antoniojimenez@gmail.com

**Abstract**

This work proposes a different procedure to encrypt images of 256 grey levels and colour, using the symmetric system Advanced Encryption Standard with a variable permutation in the first round, after the x-or operation. Variable permutation means using a different one for each input block of 128 bits. In this vein, an algorithm is constructed that defines a Bijective function between sets $N_m = \{n \; \varepsilon \; N | 0 \leq n \leq m! - 1\}$ with $m \geq 2$ and $\Pi_m = \{\pi \,|\, \pi \text{ is a permutation of } 0, 1, \ldots, m-1\}$. This algorithm calculates permutations on 128 positions with 127 known constants. The transcendental numbers are used to select the 127 constants in a pseudo-random way. The proposed encryption quality is evaluated by the following criteria: Correlation; horizontal, vertical and diagonal, Entropy and Discrete Fourier Transform. The latter uses the NIST standard 800-22. Also, a sensitivity analysis was performed in encrypted figures. Furthermore, an additional test is proposed which considers the distribution of 256 shades of the three colours; red, green and blue for colour images. On the other hand, it is important to mention that the images are encrypted without loss of information because many banking companies and some safety area countries do not allow the figures to go through a compression process with information loss. i.e., it is forbidden to use formats such as JPEG.

## 1.- Introduction.

From Internet development the problem of security sensitive information has arisen, because in this communication canal there is data theft that restricts access, such as images of areas: banking, medical, industrial, business and others. Therefore, various information encryption methods have appeared, in particular pictures (Li, J. & Gan L. 2011). There is some novel method employing the Hilbert transform (Xuemei, L., et al. 2011), Chaos (Fu, C., Chen,

J., Zou H., Meng, W., Zhan Y., & Yu, Y. 2012), the Hyper-chaos (Zhu, H., Zhao, C., & Zhang, X. 2013) or even the cryptosystem Advanced Encryption Standard- AES with a modification in the algorithm for generating keys (FIPS PUB 197. 2001).

The first three, although fast, have the problem of robustness (Douglas, & Stinson, R. 2002)., that is, the first two did not specifically mention the number of keys. In the fourth is noted the size of the set of keys, $2^{150}$, but only speak of brute force attack and differential attack (Biham, E., & Shamir, A. 1993) and the analysis is not made of some other kind of attack such as the linear (Matsui, M. 1994). In the case of AES, differential and linear attacks are not a problem, at least as presented so far (Daemen, J. & Rijmen, V. 1999), and the set of AES keys can be up to $2^{256}$ (FIPS PUB 197. 2001). Also, the AES algorithm has the substitution operation through a box. The substitution operation gives to the encryption process a nonlinearity feature (Carlet, C. 2005) that in any of the algorithms listed above does not exist except the AES algorithm. By the way, the nonlinearity of the AES box exceeds the DES and Triple-DES boxes (FIPS PUB 46-3. 1999), specifically, in the last two cryptosystems the box seven is the weakest (Silva-García, V. M. 2007).

The authors decided to use the AES algorithm for images encryption according to the following reason: because it is a recent symmetric encryption system, and besides being FIPS PUB 197 (2001) makes the AES algorithm the most studied in the world. However, an efficient method for breaking it has not yet found (Osvik, D. A., Shamir, A., & Tromer, E. 2005).

Regarding the figure encryption "quality", it is related with the randomness degree in the distribution of the encrypted image bits.

Several methods have been used to measure the randomness degree (Rukhin, A., Soto, J., Nechvatal, J., Smid, M., Barker, E., Leigh, S., Levenson, M., Vangel, M., Banks, D., Heckert, A., Dray, J., & SanVo. 2010), and those used in this research are to set out hereunder: Correlation; horizontal, vertical and diagonal; the Entropy and Discrete Fourier Transform. The latter measured the periodicity degree which has a string of zeros and ones. Furthermore, it proposes a different way to measure the randomness degree of the figure encrypted bits, using a "goodness-of-fit test" (Wolpe, R., & Myers, R. 2007).

The reason the images are not compressed is because there are some security areas in the world which do not allow compression in the information encryption process (Nom-151. 2002).

In other words, this work does not use the process: Compression - Encryption → Decryption - Decompression. Only, it would be Encryption → Decryption.

This work employs five figures ─ which appear in most encryption works─ to compare the quality of the proposed algorithm. Four are: Lena, Baboon, Pepper and Barbara. The fifth image is proposed based on a criterion presented in this research to select figures to be encrypted. Also, it is mentioned that when using a symmetric cryptosystem to encrypt images they do not all give a satisfactory result after the encryption process. This anomaly is based on the randomness degree having image bits before being encrypted, since a figure with high randomness in its bits distribution is easier to encrypt than those that have a low randomness degree in their bits (Flores-Carapia, R., Silva-García, V. M., & Rentería-Márquez, C. 2012).

This work is organized as follows: the first part presents a very synthetic state of the art. Part two shows the basic concepts used in the article. The Bijective function is addressed in section three, and the fourth section touches on the AES algorithm with variable permutation. The measure of randomness comes in part five. Point six shows the results with a sensitivity analysis, and section seven presents the conclusions.

**2.- Preliminaries.**

One algorithm long studied in the world is the AES algorithm, and for this reason it is not known yet if there is an efficient way to solve it (Osvik et al. 2005). Another important issue to clarify is that the AES cryptosystem is a symmetric algorithm, which makes it very fast to encrypt information. Moreover, when encrypting a figure that has a low randomness degree in their bits and also employs a fix key to encrypt an image, the result could be to provide information, i.e, the distribution of different tones of primary colours follows a certain pattern (Silva-García, V. M., Flores-Carapia, R., & Rentería-Márquez C. 2013), see figure 2.

This means, an additional element in the algorithm has to be included, which is a different permutation in each block of 128 bits. This permutation is applied in the first round after the x-or operation. The reason why it is not used in the entrance first round as Triple – DES (Barker, W. 2008), is because some images have areas of the same colour, such as white or black. In this situation there are strings of ones or zeros, so a permutation does not make any change in the chain. However, when it is used after the x-or operation this allows changes in the bit strings.

Why in the first round? If it is understood that the information is mixed in each round, then any changes made with the permutation in the first round will have more opportunity to scramble the information, and the zeros and ones at the end of the encryption process will appear randomly.

The transcendental numbers have the characteristic not be the solution of any polynomial, $a_n x^n + a_{n-1} x^{n-1} + \ldots a_0 = 0$ for all $a_i \varepsilon Z$ (Spivak, M. 1993), but also have the property that decimals that appear to right decimal point do not follow any periodicity, that is why they are good candidates for generating pseudo-random numbers. In fact, for this feature irrational numbers are used in the functions Has-Sha (FIPS PUB 180-3. 2008). The transcendental number used in this work is; $pi$. The reason is because it has been studied a long time (Jaohxv. 2010).

The permutation generation depends of the symmetric system key, in this case the AES key. This is based on the following criteria: the integer number associated to the 128 bits chain, that is the AES key, is denoted as $l$. Then, the product $l*pi$ is also a transcendental number so the constants can be obtained from it to generate the permutations.

Regarding the Entropy, it is measured according to the $H(x) = -\sum_{x \varepsilon X} P(x) log_2 P(x)$ and in section 5 is a wider explanation. Now if working with colour images, each basic colour — red, green or blue — is described by one byte, i.e 256 levels are sufficient and, of course, it is the same situation for mono-colour figures. Thus, if each basic colour has uniform distribution, i.e, all points are equally probable, the entropy value is 8 (Shannon, E. 1948). This means that the information is completely random. However, in practice this is not so. Thus, it will be seeking values as close as possible to 8 for each basic colour distribution.

A statistical test to evaluate the randomness of a bits sequence is formulated by a null hypothesis "$H_0$", which sets up that the bit string is random versus the alternative hypothesis "$H_a$" which indicates that this is not true. To accept or reject the null hypothesis a statistical and a threshold that define a rejection region is used. So if the statistic value based on the data, which in this case is a sequence of zeros and ones, gives a value in the rejection region this implies that the null hypothesis is not accepted. Otherwise it is accepted.

There are two types of errors in a hypothesis test, namely: type I error and type II. The type I error is one that is committed when $H_0$ is rejected, this being false. The type II error is accepting $H_0$ when this is false. The error that is controlled is the type I, because it is considered that $H_0$ is the more important of the two hypotheses.

The amount used in this research for type I error is $\alpha = 0.01$, although the value $\alpha = 0.001$ can be used (Rukhin et al. 2010).

The probability distributions that are used in this work for randomness tests are: Chi-square ($\chi^2$) and Complementary Error Function erfc (Abramowitz, M. & Stegun, I. 1964). On the other hand, erfc (z) can be expressed as a function of the cumulative standard normal distribution according to the following reasoning:

The cumulative standard normal distribution is:

1) $\Phi(z) = \frac{1}{\sqrt{2\pi}} \int_{-\infty}^{z} e^{-\frac{u^2}{2}} du$; and

(2) erfc $(z) = \frac{2}{\sqrt{\pi}} \left( \int_{z}^{\infty} e^{-u^2} du \right)$ the Complementary Error Function

The next variable change to expression $u = \frac{v}{\sqrt{2}}$ & $du = \frac{dv}{\sqrt{2}}$ is proposed

Thus, the expression (2) can be written up as follows:

erf $(z) = \frac{2}{\sqrt{2\pi}} \left( \int_{\sqrt{2}z}^{\infty} e^{-\frac{v^2}{2}} dv \right)$; if this expression is manipulated, it can be written as follows:

erfc$\left(z = \frac{z'}{\sqrt{2}}\right) = \frac{2}{\sqrt{2\pi}} \left( \int_{z'}^{\infty} e^{-\frac{v^2}{2}} dv \right)$, so

(3) erfc $\left(z = \frac{z'}{\sqrt{2}}\right) = 2 \left(1 - \frac{1}{\sqrt{2\pi}} \int_{-\infty}^{z'} e^{-\frac{v^2}{2}} dv \right) = 2(1 - \Phi(z'))$.

## 3.- The Bijective Function.

Let us start from the following facts: given a non-negative integer $m \geq 2$ can define the sets $N_m = \{n \, \varepsilon \, N | 0 \leq n \leq m! - 1\}$ and $\Pi_m = \{\pi \, | \, \pi \text{ is a permutation of } 0, 1, \ldots, m - 1\}$. Furthermore, according to the Euclidean division algorithm (Rosen, K. 2003), for all $n \, \varepsilon \, N$ it can be written in a unique way, as follows:

(4) $n = C_0(m-1)! + C_1(m-2)! + \cdots C_{m-2}(1)! + C_{m-1}(0)!$

Note that if $m$ is given, then $(m-1)!$, $(m-2)!$, ..., 1!, 0! are fixed. Moreover, it will be seen in the description of the encryption algorithm that the number $C_{m-1} = 0$. Also, it is easy to prove that:

(5) $0 \leq C_i < (m-i)$ with $0 \leq i \leq (m-2)$ (Silva, V. M., Flores, R., & Rentería C. 2009).

When the values $C_0, C_1, \ldots, C_{m-2}$, were calculated, the next algorithm can be developed:

Step 0 - An arrangement is defined in increasing order as follows: X [0] = 0, X [1] = 1, …, X [$m-1$] = $m-1$

Step 1- In accordance with the expression 5, $C_0 < (m)$, it follows that X [$C_0$] is one of the array elements in the step 0. X [$C_0$] is removed from the arrangement and its place is taken by X [$m-1$], i.e, the last element of the array. Note, in this case only two operations are performed, removal and replacement. In fact, the positions of the other array elements remain unchanged, and just the position of X [$m-1$] is reassigned.

Step 2 - Again, using the expression 5 it follows that $C_1 < (m-1)$, so X [$C_1$] is an array element in step 1. In the same way as in the previous step, X [$C_1$] is removed from the arrangement and replaced by X [$m-2$] which is the last element.

Step $m-1$ - If this process is repeated, at the end the following result will have: X [$C_{m-2}$] and X [$C_{m-1}$] = $k$ with $0 \leq k \leq m-1$. The number X [$C_{m-1}$] automatically appears because is the last element, i.e, $C_{m-1} = 0$ since there is only one position, the zero.

Finally, the positive integers array X [$C_0$], X [$C_1$],…,X [$C_{m-2}$] and X [$C_{m-1}$] is a permutation of 0, 1,…, $m-1$. This procedure is performed in $m-1$ steps.

Regarding the complexity to carry out this algorithm is $O(m)$, since in each step a removal and replacement of an element is performed, other elements remain unchanged. On the other hand, in the work "Some Cryptographic Algorithm for Strengthening Systems" (Silva et al. 2009) it is $O(m^2)$, since at each step after removing an array element, they are rearranged again. Then the number of operations performed is $(m-1) + (m-2) + \ldots + 1 = m(m-1)/2$.

It is clear in the case of images having about 1000000 pixels this means an important advantage as the complexity is only $O(m)$.

The algorithm described above defines a Bijective function (Silva et al. 2009), which is denoted by $I_m$. So, if it is Bijective it follows that it is a one to one function. This is important, since given two different positive integers $n_1, n_2$, they are associated to two different permutations, which means they modify, in general, the 128 bit block from the operation $Input \oplus k_1$ in a different way. The first schedule key and the first round entrance are $k_1$ and $Input$ respectively.

## 4.- Advanced Encryption Standard with variable permutation.

At this point the way to use the algorithm shown above to construct a permutation and how to apply this tool in the image encryption is presented. Regarding the first point, the reasoning is as follows:

Observing the developed algorithm it is easy to understand that knowing the values $C_i$ for $i = 1, 2, \ldots, m-1$, a permutation can be constructed. Note it is not important to know what the number $n$ is, fortunately, because otherwise it would work with integers of magnitude 128! - 1 $\approx 10^{215}$. In this sense, the quantities $(m-1)!, (m-2)!, \ldots$, are used as an indicator only, i.e, it is not important to write them with all their digits. The next question to address is how the values in a pseudo-random way are chosen. In this research the number *pi* is used as follows:

1) The symmetric cryptosystem key –AES- is a string of zeros and ones, which represents a positive integer, in other words, if the key length is $n + 1$ bits, it follows the integer can be written as $(a_n)2^n + (a_{n-1})2^{n-1} + \cdots + a_0$, where $a_i = 0, 1$ for $i = 0, 1, \ldots, n$. Denote this integer like $l$ and then $l$ is multiplied by *pi*, such that, the product is itself a transcendental number. Particularly in this research the AES -128 that has a 128 bits key long is used; although, it is possible to use up to 256 bits key.

2) After the multiplication is performed, it is taken to the right of the decimal point strings of one byte. In fact, for each 128 bits block 127 byte strings is necessary and the integers associated to each string is denoted as $b_0, b_1 \ldots, b_{126}$. It follows that there are as many sets of 127 bytes as 128 bit blocks have the image to be ciphered.

The constants as: $C_i = b_i$ mod. $(128 - i)$ for $i = 0, 1, \ldots, 126$; are defined remembering that the constant $C_{127} = 0$ because is the last element.

3) Once the constants, $C_i$ with $i = 0, 1, \ldots, 126$, were calculated for each 128 bits block in the image, it applied the Bijective function algorithm for each 128 bits block, which was described above. The result is the permutation of 128 positions.

If in the last procedure each set of 127 bytes is taken one after another, the required amount of these can be large. For example, for an image of 7372800 bits 57600 blocks of 128 bits are required, thus it is necessary to have 57600 * 127 = 7315200 bytes. However, this number may be reduced if the procedure is as follows: The first permutation is taken from byte 0 to 126, the second from 1 to 127, the third byte 2 to 128 and so on until the required number of permutations. Proceeding in this way the amount of bytes required in the example is, 126 + 57600 - 1 = 57725, an important reduction. Furthermore, sometimes bytes are added to the image according to the following criteria: if $24(m)$ mod. $128 \not\equiv 0$, where $m$ is the image pixel number, and then it adds the minimum number of bytes, $n$, that compliment with $24(m) + 8(n)$ mod. $128 \equiv 0$.

It is worth mentioning that the permutations generation depends of the AES-128 key. Then, in a secure communication scheme, such as Public Key Infrastructure - PKI (ElGamal, T. 1985), the key can be encrypted and transmitted using an asymmetric encryption cryptosystems like, ElGamal T. (1985), RSA (Shamir, A. & Adleman, L. 1978) or the Elliptical Curve (Koblitz, N. et al. 2000). The receiver knows the AES-128 key using the private key and then calculates $l * pi$; later she or he computes the variable permutations.

**5.- Random analysis**

*5.1- Correlation, Entropy and Discrete Fourier Transform.*

In this part the following randomness tests will be shown: Correlation; horizontal, vertical and diagonal, Entropy and Discrete Fourier Transform (Bracewell, R. N. 1986). Also, it is noted that the image encryption is performed without compression, or more specifically without information loss.

In any image encryption it is important that the distribution of the bits be random, that is, for avoiding biases that could lead to attacks by knowing the key or the original image.

Regarding the correlation between adjacent pixels of the original image, it is expected that a high correlation exists between them, that is, values close to 1 or -1 (Wolpe, R., & Myers, R. 2007). The adjacent pixels are considered in 3 directions, namely: horizontal, vertical and diagonal. Meanwhile, with a figure "well encrypted" the correlation among adjacent pixels should be an amount close to zero. The process of calculating the correlation between two random variables is performed as follows: It randomly selects an encrypted image pixel. This pixel has a level of red, green and blue which are denoted as: $x_r$, $x_g$ and $x_b$, in other words, the analysis is performed for each of the basic colours. After selecting a random pixel, the next is adjacent and it is taken in: horizontal, vertical or diagonal directions as applicable, and the same way as before, the adjacent pixel has a level of red, green and blue. These are denoted as follows: $y_r$, $y_g$ and $y_b$.

Then, it is possible to calculate the correlations in the three directions for each basic colour. The formula for calculating the horizontal direction correlation and the red colour is presented below:

(6) $r_{h;x_r,y_r} = \dfrac{\frac{1}{M}(\sum_{i=1}^{M}(x_{i,r} - \bar{x}_r)(y_{i,r} - \bar{y}_r))}{\sqrt{\left(\frac{1}{M}\sum_{i=1}^{M}(x_{i,r} - \bar{x}_r)^2\right)\left(\frac{1}{M}\sum_{i=1}^{M}(y_{i,r} - \bar{y}_r)^2\right)}}$ where $\bar{x}_r$, $\bar{y}_r$ can be expressed as:

(7) $\bar{x}_r = \frac{1}{M}\sum_{i=1}^{M} x_{i,r}$ y $\bar{y}_r = \frac{1}{M}\sum_{i=1}^{M} y_{i,r}$

Clearly, the formulas are the same in other directions and colours.

The Entropy analysis takes into account the pixels dispersion of the encryption images. This analysis is carried out by separating the basic colours. In case of a single colour only one byte to represent the Entropy is required, that is, 256 levels for basic colours. In this vein, if the distribution basic colour is completely random then the Entropy is 8 for each one, but in practical cases, the randomness measured in a string of zeros and ones have the following reasoning: when the Entropy value is close to 8 it is understood the ones and zeros chain is random, otherwise, it means that it is not.

To calculate the Entropy it is assumed that there is a pixels chain. In this sense, it is possible to separate the pixels chain in the primary colours. Then assuming that the bit string corresponds to the red colour which in turn can be divided into 8 bit blocks, that is, one byte, it follows there are 256 possible values. Their frequencies are recorded in a table of 256 classes according to their appearance. Then, each class is assigned a frequency $f_i$ for $i = 0, 2...,$ 255, so that a probability estimate of the class occurrence is $P_i[x_i] = 1/f_i$, where, $f_i$ is the $x_i$ frequency, with $i = 0, 2 ..., 255$. Thus, the Entropy for the red colour is calculated as follows: $H_r = - \sum_{x_i \varepsilon X} P_r(x_i) log_2 P_r(x_i)$, where $X$ is the set of all 256 values. In a simple way, it is easy to see that the expressions for the green and blue colours are the same.

The Discrete Fourier Transform measured the randomness degree of zeros and ones string, i.e. there are no periodicity repetitive patterns one after another in the occurrence of zeros and ones.

In addition, the following elements are involved in the calculation of the statistic test:

The number $N_0$ is the expected theoretical amount, $(0.95)\ n/2$. Where, the chain length is $n$.

The number $N_1$ is the amount of values less than a dimension $h$, which depends on the $n$ [24].

$f_j = \sum_{k=1}^{n} x_k e^{[2\pi i(k-1)j/n]}$. Where $i = \sqrt{-1}$ and $j = 1, 2 \cdots n/2 - 1$. If the $n$ is odd, it simply removes the last bit of the string. Clearly, the $f_j$ has a real and complex part. Then, the modulus $\|f_j\|$ is calculated, which is compared with $h$. If the $\|f_j\| < h$ a one is added to $N_1$ value. Otherwise, $N_1$ stays the same value.

With this data it is possible to calculate the variable $d = \frac{N_1 - N_0}{\sqrt{n(.95)(0.05)/4}}$ and the test statistic $P - value = erfc(\frac{d}{\sqrt{2}})$;

and $erfc(\frac{d}{\sqrt{2}}) = 2(1 - \Phi(d))$. The decision rule says: if $P - value$ is less than 0.01 the null hypothesis is rejected, otherwise it is accepted. The three tests are illustrated in Section 6 for a particular 128 bit key value.

*5.2.- Randomness test proposal.*

When working with images a randomness test is added, this is based on the manner in which the pixels are arranged in an encrypted figure. In this sense, the statistics $\chi^2 = \sum_{i=1}^{i=k} \left[\frac{(o_i - e_i)^2}{e_i}\right]$ is proposed to be used for each basic colour, where $o_i$ is the observed value and $e_i$ is the expected amount. Using this statistic it is possible to quantify the randomness degree that has the shades distribution of the basic colours, red, green and blue. All NIST 800-22 tests do not use this type of test when it is necessary to know the randomness degree of a bit string, i.e. they do not measure the randomness of the tones distribution in the basic colours, red, green and blue of the image encrypted.

In the same way as some NIST 800-22 standard tests, this proposal applies the goodness-of-fit test using the statistic, $\chi^2$, which has a probability distribution Chi-square. This distribution has $n - 1$ degrees of freedom. The freedom degrees are obtained in the following way: the shades of each image colour can be displayed as a histogram, whose abscissa has 256 divisions. Thus, the freedom degrees are 256 – 1= 255 (Nom-151. 2002). Furthermore, considering the random variable $\chi^2$ approaches to normal distribution according to the Central Limit Theorem (Bracewell, R. N. 1986), it follows that; the statistic $\chi^2$ has a mean and standard deviation:
μ = 255 and σ = [2 (255)]$^{1/2}$ = 22.5831.

With this information it is simple to calculate the thresholds for significance level α = 0.01 and α = 0.001, taking into account that both thresholds are in the normal distribution right tail. So, the threshold for significance level α = 0.01 is 307.61 and for α = 0.001 is 324.78, then, the decision process to accept or reject the null hypothesis according to a particular data chain is as follows:

a) Calculating the statistic $\chi^2 = \sum_{i=1}^{i=k} \left[\frac{(o_i - e_i)^2}{e_i}\right]$ for specific values $o_i$ and $e_i$; and as before these are considered the $i$ −value for the observed and expected amounts.

b) The probability that there is to right of $\chi^2$ particular value is calculated. If this probability is greater than or equal to 0.01, then the null hypothesis is accepted, otherwise it is rejected. Of course, in the case of 0.001 the procedure is in a similarly way.

*5.3.- Criteria for selecting the image*

In the introduction to this article it was mentioned that a criterion for choosing the figure to be encrypted would be shown. This criterion is based on a characteristic of the goodness-of-fit test that tells the following: if the distribution of tones in each of the three basic colours were totally random $\chi^2 = 0$. In fact, this means the colours histogram is a uniform distribution.

However, if the colours have a very large value it means they follow a defined order. So, this research proposes to choose an image that has a large as possible $\chi^2$ value for each colour. Another reason is because in many images with a "small" $\chi^2$ for each colour, for example, less than half a million, the AES cryptosystem can be applied directly, i.e., it is not necessary to use a random permutation to cipher an image that approves the randomness tests presented above, and even the proposed test. This is shown in Figure 3. On the other hand, figures with very large $\chi^2$ values for the basic colours, for example, more than one hundred million, it needs to apply variable permutations in order to encrypt an image that passes the randomness tests. This case is presented in Figure 2.

Hombres necios que acusáis
a la mujer sin razón,
sin ver que sois la ocasión
de lo mismo que culpáis:
si con ansia sin igual
solicitáis su desdén,
por qué queréis que obren bien
si las incitáis al mal?
Combatís su resistencia
y luego, con gravedad,
decís que fue liviandad
lo que hizo la diligencia.
Parecer quiere el denuedo
de vuestro parecer loco,
al niño que pone el coco
y luego le tiene miedo.

Figure 1.- The proposed image to be ciphered with the next chi-square for basic colours $\chi_r^2 = 191616040.31$, $\chi_g^2 = 193547200.97$ y $\chi_b^2 = 190551369.15$.

In this paper the figure proposed has the next chi-square amounts for each basic colour red, green and blue: $\chi_r^2 = 191616040.31$, $\chi_g^2 = 193547200.97$ and $\chi_b^2 = 190551369.15$ respectively. This image corresponds to a stanza of poem written by Sor Juana Ines de la Cruz (Mexican Poetess 1615-1683) (Paz, O. 2012), which is shown in figure 1. When figure 1 is encrypted using AES cryptosystem without variable permutations, the cipher image, at least, gives some information from the original text. This is shown in figure 2.

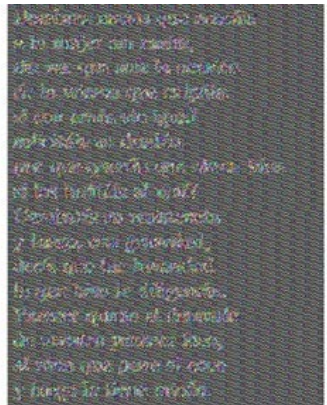

Figure 2.- Figure 1 encrypted without variable permutations.

Furthermore, the figure 4 image c) has the following chi-square for each primary colour red, green and blue: $\chi_r^2 = 113272.3$, $\chi_g^2 = 247937.56$ and $\chi_b^2 = 344545.33$ respectively. The result of the figure 4 image c) encryption without applying variable permutations is presented in Figure 3. This is basically the argument why the proposed image has $\chi^2$ as largest as possible in each basic colour, in order to verify that the cryptosystem presented here is efficient; that is, the encrypted figure approves all the randomness tests.

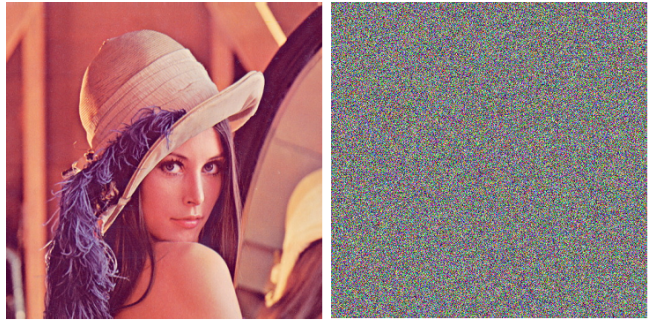

Figure 3.- The figure 4 image c) encrypted without variable permutations.

**6.- Result Presentation**

This section presents the results when the tests already mentioned are applied to five images: those are in figure 4 and 1. The 128-bit key chosen for AES cryptosystem is written in hexadecimal system as follows: $k$ =0123456789ABCDEFFEDCBA9876543210. In fact, it can be assigned arbitrarily.

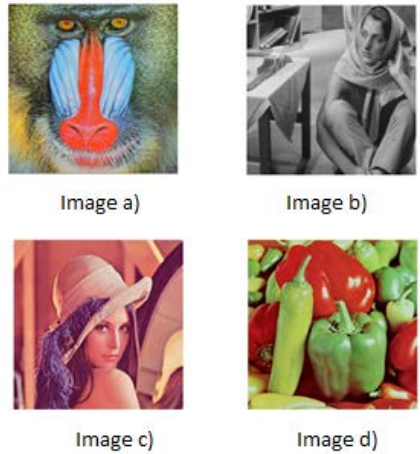

Figure 4- Image a) Baboon, b) Barbara, c) Lena and d) Peppers.

As noted before, the key $k$ is associated with a positive integer, which is presented next: $l = 94522879700260684207971970630038032$. Later, this number is multiplied by $pi$ and the product $l * pi$ is a transcendental number too. After this, it is taken to the right of decimal point the number of bits needed to cover all the constants sets used to encrypt the image.

In another vein, the analysis in this research is carried out in colour figures, but can also be encrypted figures of 256 levels of grey, as is the case of figure 5. However, both kinds of images give as a result a colour encrypted image.

*6.1.- Results of the Discrete Fourier Transform and the test proposed.*

This applies the Discrete Fourier Transform and the proposed test to images encrypted to find out the randomness. Figure 4 and 1 are used with the $k$ key to verify if the chains of each basic colour: red, green and blue approve the randomness criteria. The results are presented in Table 1.

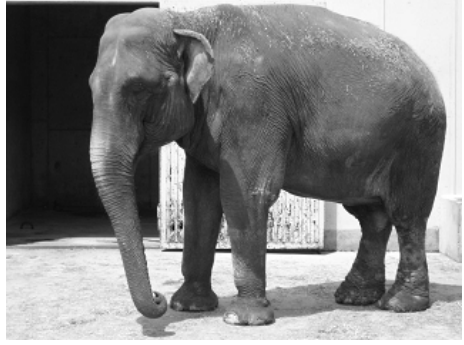
Figure 5.- Image with 256 grey levels.

*6.2.- Correlation analysis between adjacent pixels.*
It is expected that in the original image the adjacent pixels in horizontal, vertical or diagonal directions, have correlations close to 1 or -1 (Wolpe, R., & Myers, R. 2007). However, for adjacent pixels in the same three directions but in an encrypted image, it is expected that their values are close to zero. For calculating the correlation a random sample of 3,000 adjacent pixels pairs from the image was taken. The results of the correlations for the original images are shown in tables 2, 3 and 4. The encrypted images are presented in table 5, 6 and 7.

Table 1.- Randomness results of the figure 4 and 1 cipher images.

| Test name | Significance Label | Image a) $P-value$ and Decision | Image b) $P-value$ and Decision | Image c) $P-value$ and Decision | Image d) $P-value$ and Decision | Figure 1 $P-value$ and Decision |
|---|---|---|---|---|---|---|
| Spectral DTF test | α = .01 | $P-value_r = 0.31$; Accepted | $P-value_r = 0.25$; Accepted | $P-value_r = 0.76$; Accepted | $P-value_r = 0.50$; Accepted | $P-value_r = 0.47$; Accepted |
|  |  | $P-value_g = 0.67$; Accepted | $P-value_g = 0.63$; Accepted | $P-value_g = 0.63$; Accepted | $P-value_g = 0.30$; Accepted | $P-value_g = 0.99$; Accepted |
|  |  | $P-value_b = 0.54$; Accepted | $P-value_b = 0.21$; Accepted | $P-value_b = 0.97$; Accepted | $P-value_b = 0.73$; Accepted | $P-value_b = 0.36$; Accepted |
| The Proposed Test | α = .01 | $P-value_r = 0.03$; Accepted | $P-value_r = 0.25$; Accepted | $P-value_r = 0.77$; Accepted | $P-value_r = 0.65$; Accepted | $P-value_r = 0.27$; Accepted |
|  |  | $P-value_g = 0.56$; Accepted | $P-value_g = 0.12$; Accepted | $P-value_g = 0.25$; Accepted | $P-value_g = 0.14$; Accepted | $P-value_g = 0.45$; Accepted |
|  |  | $P-value_b = 0.66$; Accepted | $P-value_b = 0.15$; Accepted | $P-value_b = 0.81$; Accepted | $P-value_b = 0.77$; Accepted | $P-value_b = 0.06$; Accepted |

Table 2.- The correlation results in horizontal, vertical and diagonal directions for red colour using images of figures 4 and 1.

| Test Name | Image a) | Image b) | Image c) | Image d) | Figure 1 |
|---|---|---|---|---|---|
| Correlation of red colour | $r_{h;r}= 0.9076$ | $r_{h;r}= 0.9010$ | $r_{h;r}= 0.9760$ | $r_{h;r}= 0.9789$ | $r_{h;r}= 0.8080$ |
| | $r_{v;r}= 0.8637$ | $r_{v;r}= 0.9598$ | $r_{v;r}= 0.9871$ | $r_{v;r}= 0.9795$ | $r_{v;r}= 0.8253$ |
| | $r_{d;r}= 0.8332$ | $r_{d;r}= 0.8716$ | $r_{d;r}= 0.9613$ | $r_{d;r}= 0.9696$ | $r_{d;r}= 0.6971$ |

Table 3.- The correlation results in horizontal, vertical and diagonal directions for green colour using images of figures 4 and 1.

| Test Name | Image a) | Image b) | Image c) | Image d) | Figure 1 |
|---|---|---|---|---|---|
| Correlation of green colour | $r_{h;a}= 0.8579$ | $r_{h;a}= 0.9010$ | $r_{h;a}= 0.9742$ | $r_{h;a}= 0.9897$ | $r_{h;a}= 0.8084$ |
| | $r_{v;a}= 0.7780$ | $r_{v;a}= 0.9598$ | $r_{v;a}= 0.9850$ | $r_{v;a}= 0.9906$ | $r_{v;a}= 0.8287$ |
| | $r_{d;a}= 0.7289$ | $r_{d;a}= 0.8716$ | $r_{d;a}= 0.9586$ | $r_{d;a}= 0.9836$ | $r_{d;a}= 0.6898$ |

Table 4.- The correlation results in horizontal, vertical and diagonal directions for blue colour using images of figures 4 and 1.

| Test Name | Image a) | Image b) | Image c) | Image d) | Figure 1 |
|---|---|---|---|---|---|
| Correlation of blue colour | $r_{h;b}= 0.9185$ | $r_{h;b}= 0.9010$ | $r_{h;b}= 0.9524$ | $r_{h;b}= 0.9760$ | $r_{h;b}= 0.8145$ |
| | $r_{v;b}= 0.8801$ | $r_{v;b}= 0.9598$ | $r_{v;b}= 0.9705$ | $r_{v;b}= 0.9786$ | $r_{v;b}= 0.8316$ |
| | $r_{d;b}= 0.8578$ | $r_{d;b}= 0.8716$ | $r_{d;b} = 0.9252$ | $r_{d;b}= 0.9614$ | $r_{d;b}=0.6979$ |

Table 5.- The correlation results in horizontal, vertical and diagonal directions for red colour using cipher images of figures 4 and 1.

| Test Name | Cipher image a) | Cipher image b) | Cipher image c) | Cipher image d) | Cipher figure 1 |
|---|---|---|---|---|---|
| Red colour correlation | $r_{h;r}= 0.0092$ | $r_{h;r}= 0.0023$ | $r_{h;r}= 0.0070$ | $r_{h;r}= 0.0152$ | $r_{h;r}= 0.0067$ |
| | $r_{v;r}= 0.0034$ | $r_{v;r}= 0.0143$ | $r_{v;r}= 0.0138$ | $r_{v;r}= 0.0114$ | $r_{v;r}= 0.0259$ |
| | $r_{d;r}= 0.0248$ | $r_{d;r}= 0.0028$ | $r_{d;r}= 0.0088$ | $r_{d;r}= 0.0119$ | $r_{d;r}= 0.0026$ |

Table 6.- The correlation results in horizontal, vertical and diagonal directions for green colour using cipher images of figures 4 and 1.

| Test Name | Cipher image a) | Cipher image b) | Cipher image c) | Cipher image d) | Cipher figure 1 |
|---|---|---|---|---|---|
| Green colour correlation | $r_{h;a}= 0.0154$ | $r_{h;a}= 0.0034$ | $r_{h;a}= 0.0176$ | $r_{h;a}= 0.0259$ | $r_{h;a}= 0.0023$ |
| | $r_{v;a}= 0.0078$ | $r_{v;a}= 0.0175$ | $r_{v;a}= 0.0060$ | $r_{v;a}= 0.0066$ | $r_{v;a}= 0.0239$ |
| | $r_{d;a}= 0.0217$ | $r_{d;a}= 0.0015$ | $r_{d;a}= 0.0165$ | $r_{d;a}= 0.0119$ | $r_{d;a}= 0.0219$ |

Table 7.- The correlation results in horizontal, vertical and diagonal directions for blue colour using cipher images of figures 4 and 1.

| Test Name | Cipher image a) | Cipher image b) | Cipher image c) | Cipher image d) | Cipher figure 1 |
|---|---|---|---|---|---|
| Blue colour correlation | $r_{h;b}= 0.0197$ | $r_{h;b}= 0.0112$ | $r_{h;b}= 0.0244$ | $r_{h;b}= 0.0176$ | $r_{h;b}= 0.0080$ |
| | $r_{v;b}= 0.0063$ | $r_{v;b}= 0.0169$ | $r_{v;b}= 0.0077$ | $r_{v;b}= 0.0115$ | $r_{v;b}= 0.0037$ |
| | $r_{d;b}= 0.0046$ | $r_{d;b}= 0.0018$ | $r_{d;b}= 0.0088$ | $r_{d;b}= 0.0193$ | $r_{d;b}= 0.0126$ |

It is considered that the advantage of this work is to use AES, with some innovations. Besides, the set keys can reach $2^{256}$ elements. Furthermore, it is an algorithm widely studied (FIPS PUB 197. 2001) and as yet the way to resolve has not been found. On the other hand, the graphs of the basic colours distribution for both the original and its corresponding encrypted image are presented in figure 6. In this research, only for the image c) figure 4 shows the correlation horizontal, vertical and diagonal of the basic colours. They appear from up to down in figure 6. In other words, in the first line the horizontal correlation appears, and in the second the vertical and in the third the diagonal.

Furthermore, the first, third and fifth columns of the graph present the correlation in the three directions of the original image. In the other columns the three basic colour distributions of the encrypted image are shown.

*6.3.- Analysis of Entropy.*
Regarding the entropy, the randomness analysis is carried out for the three basic colours, employing both the original and the encrypted images. If a figure was "well encrypted" it is considered that the entropy has to be very close to 8. The results for the images of figure 4 and 1 are reported in Table 8.

Table 8.- The entropy results for the images encrypted for the figures 4 and 1, using $k$.

|  | Image a) | Image b) | Image c) | Image d) | Figure 1 |
|---|---|---|---|---|---|
| Red colour Entropy | 7.99917 | 7.99925 | 7.99934 | 7.99932 | 7.99978 |
| Green colour Entropy | 7.99930 | 7.99922 | 7.99925 | 7.99923 | 7.99979 |
| Blue colour Entropy | 7.99932 | 7.99923 | 7.99935 | 7.99934 | 7.99977 |

The results reported in Table 8 are better than those shown in other studies (Zhu et al. 2013).

Table 9.- The correlation between encrypted figures of image c) figure 4, for the $k$ and $k+1$ keys.

| Test Name | Sensitivity analysis for image c) using $k$ and $k+1$ keys |
|---|---|
| Red colour correlation | 0.021 |
| Green colour correlation | 0.033 |
| Blue colour correlation | 0.005 |

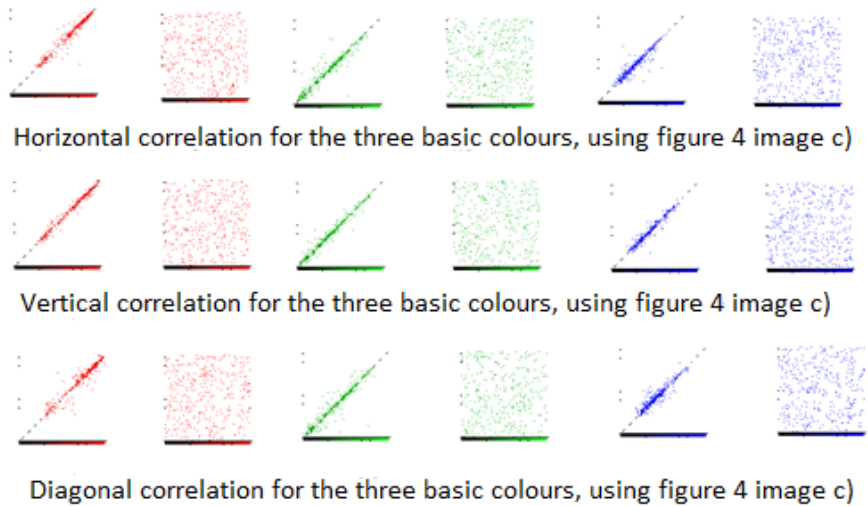

Horizontal correlation for the three basic colours, using figure 4 image c)

Vertical correlation for the three basic colours, using figure 4 image c)

Diagonal correlation for the three basic colours, using figure 4 image c)

Figure 6.-From up to down the correlations: horizontal, vertical and diagonal are shown. The first, third and fifth columns are correlations for the original image, the other corresponds to the encrypted image.

*6.4.- Sensitivity analysis.*
It is important in encryption images to carry out the correlation between two figures encrypted of the same image and with two keys very close, for example, the difference between both keys could be one. On the other hand, it is hoped that the correlation between such cipher figures should be near to zero. This would mean that no matter the closeness between different keys, there is no relationship between the two encrypted figures. As illustration, the key $k$ and $k + 1$ applied to the figure 4 image c) is considered. Thus 3000 pixel pairs are chosen where a first pixel is chosen at random in the encrypted image with $k$ and the other pixel is taken in the same position as the first but in the encrypted image with $k + 1$. The correlation analysis is performed for the three basic colours, red, green and blue. The results are presented in table 9.

**7.- Conclusions.**
This research proposes a different procedure to encrypt colour images using the AES cryptographic system with variable permutations.

Furthermore, it is pointed out that it is possible to encrypt mono-colour figures with the same procedure, i.e, those that only have 256 shades of grey. However, the encrypted image is a multi colour figure. If the randomness tests are separated in two; then, the first can be reported as: the Discrete Fourier Transform and the method proposed by the authors. This last analyzes the encrypted image strings for each colour. Besides, this procedure uses a statistic that has a $\chi^2$ distribution with 255 freedom degrees. Furthermore, both procedures use a hypothesis test of goodness-of-fit.

The second option for the randomness measurement is: the correlation and entropy. Correlation is assessed in three directions, namely: horizontal, vertical and diagonal. This procedure is used only for the image c) figure 4, with a key which can be chosen arbitrarily. The result graph for both the original and encrypted images are reported in figure 6.

Regarding the Entropy results that appear in table 8, they were compared with other work (Zhu et al. 2013), improving the information disorder for the encrypted images with the procedure proposed in this research.

On the other hand, in table 9 a sensitivity analysis is reported, regarding $k$ and $k + 1$ keys. The result tells that there is no relationship between the images encrypted with $k$ and $k + 1$.

Another important issue to note is the encryption time consumed when using a program developed in C + + builder. For image 512 x 512 pixels it is around 1.5 seconds.

Regarding the robustness of the AES cryptosystem with the proposed innovation, it can reach $2^{256}$ the set keys. Also, as AES is the FIPS PUB 197 it is one of the most studied algorithms in the world, and still has not been resolved.

**Acknowledgements** The authors would like to thank the Instituto Politécnico Nacional (Secretaría Académica, COFAA, SIP, CIDETEC, ESCOM, and ESFM), the CONACyT, and SNI for their economic support to develop this work.

**References**

Abramowitz, M. & Stegun, I. (1964). Handbook of mathematical functions. *Applied Mathematics Series. Vol. 55*, Washington: National Bureau of Standards, 1964, reprinted 1968 by Dover Publications, New York.

Barker, W. (2008). Recommendation for the Triple Data Encryption Algorithm (TDEA) block cipher. *NIST Special Publication* 800-67.

Biham, E., & Shamir, A. (1993). Differential cryptanalysis of the full 16-round DES. *Lecturer Notes in computer Science*, pp. 494-502.

Bracewell, R. N. (1986). *The Fourier Transform and Its Applications*. New York: McGraw-Hill.

Carlet, C. (2005). On highly nonlinear S-boxes and their inability to thwart DPA attacks. *6th International Conference on Cryptology of the Springer-Verlag*, pp. 49-62.

Daemen, J. & Rijmen, V. (1999). AES proposal: Rijndael, AES algorithm Submission.

Douglas, & Stinson, R. (2002). *CRYPTOGRAPHY: Theory and practice*. Chapman & Hall/ CRC Press, pp. 161-280.

ElGamal, T. (1985). A public key cryptosystem and a signature scheme based on discrete logarithms. *IEEE Transactions on Information Theory*, 469-472.

FIPS PUB 180-3. (2008). Federal Information Processing Standards Publications 180-3.

FIPS PUB 197. (2001).   Federal Information Processing Standards Publications 197.

FIPS PUB 46-3. (1999). Federal Information Processing Standards Publication 46-3.

Flores-Carapia, R., Silva-García, V. M., & Rentería-Márquez, C. (2012). DES Block of 96 bits: an application to image encryption. *International Journal of Research & Reviews in Applied Sciences*, Vol. 14.

Fu, C., Chen, J., Zou H., Meng, W., Zhan Y., & Yu, Y. (2012). A chaos-based digital image encryption scheme with an improved diffusion strategy. *Optics Express*, Vol. 20.

Jaohxv. (2010). Pi World. ja0hxv.calico.jp/pai/estart.html.

Koblitz, N. et al. (2000). Designs codes and cryptography. *The state of elliptic curve,* Vol. 19, pp. 173-193.

Li, J. & Gan L. (2011). Study on chaotic cryptosystem for digital image encryption. *Third International Conference Measuring Technology and Mechatronics Automation, IEEE*, pp. 426-430.

Matsui, M. (1994). Linear cryptanalysis for DES cipher, *Lecture Notes in Computer Science*, pp. 386-397.

Nom-151. (2002). Norma Oficial Mexicana NOM – 151 – SCFI – 2002, Prácticas comerciales – Requisitos que deben observarse para la conservación de mensajes de datos.

Osvik, D. A., Shamir, A., & Tromer, E. (2005). Cache Attacks and Countermeasures: the Case of AES. Extended Version, pp. 11-20.

Paz, O. (2012). *Sor Juana Inés de la Cruz: las trampas de la fe*. Fondo de Cultura Económica, Décimo segunda reimpresión, México.

Rosen, K. (2003). *Discrete Mathematics and its Applications*. Mc. Graw Hill 4fth edition, USA, pp. 177-180.

Rukhin, A., Soto, J., Nechvatal, J., Smid, M., Barker, E., Leigh, S., Levenson, M., Vangel, M., Banks, D., Heckert, A., Dray, J., & SanVo. (2010). A statistical test suite for random and pseudorandom number generators for cryptographic applications. National Institute of Standards and Technology, NIST 800-22.

Shamir, A. & Adleman, L. (1978). A method for obtaining digital signatures and public key cryptosystems. *Communications of the ACM*, pp. 120-126.

Shannon, E. (1948). A mathematical theory of communication. *Bell Systems Technical Journal*, 27, 377-423, 623-656.

Silva, V. M., Flores, R., & Rentería C. (2009). Algorithm for strengthening some cryptography systems. *Journal of applied Mathematics and Decision Sciences*, Hikari, pp. 967-976.

Silva-García, V. M. (2007). Criptoanálisis para la modificación de los estándares DES y Triple DES, Tesis Doctoral, CIC-IPN.

Silva-García, V. M., Flores-Carapia, R., & Rentería-Márquez C. (2013). Triple-DES block of 96 bits: An application to colour image encryption. *Applied Mathematical Sciences*, Hikari, Vol. 7, no. 23, pp. 1143–1155.

Spivak, M. (1993). *Calculus: Cálculo Infinitesimal*. Reverte, Barcelona España.

Wolpe, R., & Myers, R. (2007). *Probability and statistics for engineers and scientists*. Prentice Hall.

Xuemei, L., et al. (2011). A novel scheme reality preserving image encryption. *Third International Conference Measuring Technology and Mechatronics Automation, IEEE*, 2011, pp. 218-221.

Zhu, H., Zhao, C., & Zhang, X. (2013). A novel image encryption?compression scheme using hyper chaos and Chinese remainder theorem. *Elsevier, Signal processing: Image Communication*, pp. 670-680.